\documentclass[letterpaper,letter,dvipdfmx]{jpsj3}
\usepackage{graphicx}
\usepackage{times}
\usepackage{color}
\usepackage{amsmath,amssymb,bm}
\usepackage{here}

\title{Unveiling Quadrupolar Kondo Effect in the Heavy Fermion Superconductor PrV$_2$Al$_{20}$}

\author{Mingxuan Fu$^1$, Akito Sakai$^1$, Naoki Sogabe$^1$, Masaki Tsujimoto$^1$, Yosuke Matsumoto$^2$, \\ and Satoru Nakatsuji$^1$$^3$$^4$$^5$\thanks{E-mail address: satoru@phys.s.u-tokyo.ac.jp}}

\inst{$^1$Institute for Solid State Physics, University of Tokyo, Kashiwa, Chiba 277-8581, Japan \\
$^2$Department of Quantum Materials, Max Planck Institute for Solid State Research, Heisenbergstrasse 1, Stuttgart 70569, Germany \\
$^3$Department of Physics, Faculty of Science and Graduate School of Science, The University of Tokyo \\
$^4$CREST, Japan Science and Technology Agency (JST), 4-1-8 Honcho Kawaguchi, Saitama 332-0012, Japan \\
$^5$ Institute for Quantum Matter and Department of Physics and Astronomy, Johns Hopkins University, Baltimore, MD 21218} 

\abst{Heavy fermion metals hosting multipolar local moments set a new stage for exploring exotic spin-orbital entangled quantum phases. In such systems, the quadrupolar Kondo effect serves as the key ingredient in the orbital-driven non-Fermi liquid (NFL) behavior and quantum critical phenomena. The cubic heavy fermion superconductor Pr$Tr_2$Al$_{20}$ ($Tr$ : Ti, V) is a prime candidate for realizing the quadrupolar Kondo lattice. Here, we present a systematic study of the NFL phenomena in PrV$_2$Al$_{20}$ based on magnetoresistance (MR), magnetic susceptibility and specific heat measurements. Upon entering the NFL regime, we observe a universal scaling behavior expected for the quadrupolar Kondo lattice in PrV$_2$Al$_{20}$, which indicates a prominent role of the quadrupolar Kondo effect in driving the NFL behavior. Deviations from this scaling relation occur below $\sim 8$ K, accompanied by a sign change in the MR and a power-law divergence in the specific heat. This anomalous low-temperature state points to the presence of other mechanisms which are not included in the theory, such as heavy fermion coherence or multipolar quantum critical fluctuations.}

\kword{PrV$_2$Al$_{20}$, quadrupolar Kondo lattice, non-Fermi liquid behavior, nonmagnetic ground-state doublet}

\begin{document}
\maketitle

The concept of quasiparticles lies at the heart of Landau's Fermi liquid theory and plays a central role in the traditional paradigms of condensed matter physics. One major mystery in strongly correlated materials is the non-Fermi-liquid (NFL) or strange metal phase featuring a complete breakdown of the quasiparticle picture and singular behavior of physical properties. Understanding of the NFL may hold the key to elucidating the driving mechanism of unconventional superconductivity and other emergent quantum states in itinerant electron systems\cite{Keimer2015,Si2016,Stewart2001, Gegenwart2008,Lohneysen2007}. Some prototypical examples of the NFL occur in heavy-fermion metals\cite{Yuan2104,Custers2003,Nakatsuji2008,Matsumoto2011,Tomita506}. In the conventional view of heavy fermion systems, the competition between the magnetic Kondo effect and the Ruderman-Kittel-Kasuya-Yosida (RKKY) interaction among local dipolar moments renders the NFL behavior, which is often attributed to magnetic quantum criticality\cite{Gegenwart2008,Lohneysen2007,Coleman_2001}. 

The $f$-electron systems hold the possibility of harboring a nonmagnetic crystal electric field (CEF) ground state in which magnetic dipolar moment is absent but higher-order multipolar moments are active\cite{KuramotoReview2009, Sakai2011,OnimaruReview2016}. The conventional picture of the Kondo effect is dramatically altered in such a system. Theoretical works on quadrupolar systems suggest that the Kondo entanglement would take place between the local quadrupolar moments and the orbital degrees of freedom of the conduction ($c$) electrons, while the $c$-spins simply offer two separated scattering channels. The resulting quadrupolar Kondo effect may open the window into new types of NFL states without magnetic critically\cite{Cox1987, Cox1998, Tsuruta2015,Adarsh2019}. 

The cubic heavy-fermion materials Pr$Tr_2$Al$_{20}$ ($Tr$ : Ti, V) provide an ideal platform for realizing quadrupolar Kondo effect\cite{Sakai2011}. The nonmagnetic  $\Gamma_3$ ground state of these compounds carries both electric quadrupoles and magnetic octupoles. In their crystal structures, the localized Pr moment resides at the center of the Frank-Kasper cage involving sixteen Al atoms\cite{Kangas2012}; this highest possible coordination number results in strong hybridization effects between the local multipolar moments and the conduction electrons \cite{Sakai2011}, as demonstrated by the Kondo resonance peak observed through photoemission spectroscopy\cite{Matsunami2011}. As a result, the substantial Kondo coupling between the local multipolar moments and the conduction electrons give rise to a rich phase diagram comprising both multipolar order and exotic superconductivity \cite{Sakai2011,Matsubayashi2014,Tsujimoto2014a,Sato2012, Ito2011, Ito2015, Taniguchi2016, Nakanishi2018}. 


In contrast to PrTi$_2$Al$_{20}$ which displays ordinary Fermi-liquid-like behavior above its ferroquadrupolar ordering temperature at ambient pressure, PrV$_2$Al$_{20}$ exhibits a NFL electronic state above the antiferroquadrupolar ordering temperature $T_{Q}$, with $\sim \sqrt T$ dependent resistivity and $\sim -\sqrt T$ Van Vleck susceptibility\cite{Sakai2011}. The reason behind this distinction is the stronger $c$-$f$ hybridization effects in PrV$_2$Al$_{20}$, as evident from the enhanced effective mass, hyperfine constant, and Seebeck effect\cite{Tsujimoto2014a,Tokunaga2013, Machida2015}. Moreover, it exhibits heavy fermion superconductivity at $T_{\rm c}$ = 0.05 K in the antiferroquadrupolar phase\cite{Tsujimoto2014a}. Magnetoresistivity measurements of PrV$_2$Al$_{20}$ under a [111] magnetic field reveal a field-induced quantum critical point upon full suppression of the long-range multipolar order, suggesting the vital role of the multipolar quantum fluctuations in shaping its phase diagram\cite{Shimura2015}.  Although the quadrupolar Kondo effect has been proposed to be the origin of the observed phenomena, conclusive evidence is still lacking. Therefore, a direct comparison between experiments and theory is crucial for clarifying the nature of the non-Fermi liquid phase and the novel quantum critical behavior in PrV$_2$Al$_{20}$. 

In this letter, we report on the magnetic and quadrupolar Kondo effect in PrV$_2$Al$_{20}$ and discuss the origin of its NFL behaviors. The resistivity shows a logarithmic increase at high temperatures, stemming from the presence of the conventional magnetic Kondo effect due to the excited multiplets. Upon further cooling, we found that the NFL state in PrV$_2$Al$_{20}$ is well coupled with generic scaling behaviors, namely, a generalized Schlottmann's scaling relation for magnetoresistance (MR) \cite{Anders1997} and the scaling relation demanded by the quadrupolar Kondo lattice model \cite{Tsuruta2015}. This finding indicates the crossover to a regime dominated by the quadrupolar Kondo effect. In this regime, the quadrupolar moments are highly entangled with the conduction electrons, triggering the observed NFL phase. 

Single crystals of PrV$_2$Al$_{20}$ and LaV$_2$Al$_{20}$ were synthesized by the Al self-flux method under vacuum or in an inert argon atmosphere at ambient pressure, as reported in earlier studies\cite{Sakai2011}. The zero-field residual resistivity ratio (RRR) of the PrV$_2$Al$_{20}$ and LaV$_2$Al$_{20}$ samples are 7 and 12 respectively, indicating their relatively high quality. We conducted magnetoresistivity measurements over a wide temperature range of 2 K to 80 K with a maximum magnetic field of 8 T using a standard four-probe ac method. To minimize the contribution from classical magnetoresistance induced by the Lorentz force, we focus on the longitudinal magnetoresistance, with magnetic field oriented parallel to the electrical current. The dc magnetic susceptibility was measured by a commercial SQUID magnetometer. For all results represented below, the applied magnetic field lies along the [110] direction. 

\begin{figure}[t]
\begin{center}
\includegraphics[keepaspectratio, scale=0.43]{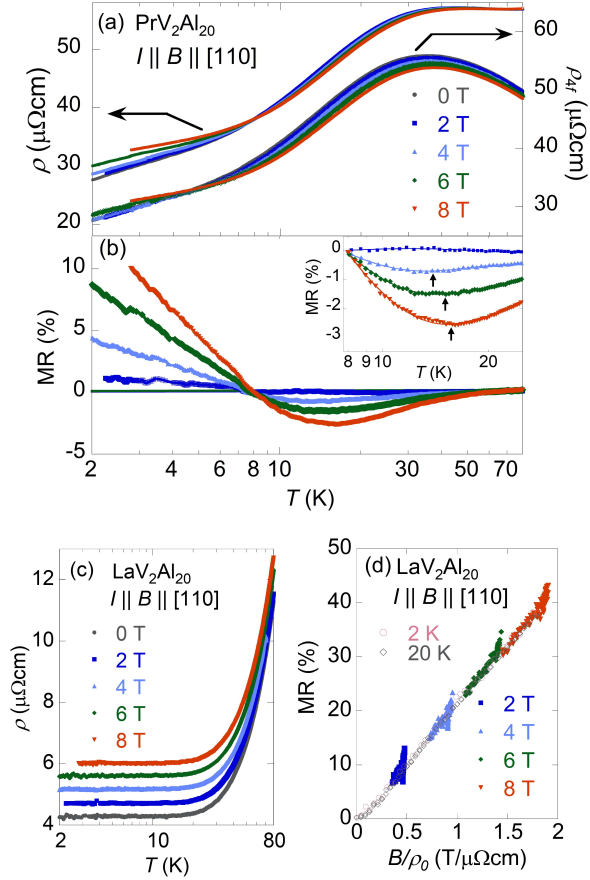}
\end{center}
\caption{(Color online) (a) Temperature dependence of the electric resistivity $\rho(T)$ (left axis) and $4f$-electron contribution $\rho_{4f}(T)$ (right axis) of PrV$_2$Al$_{20}$ under various magnetic fields along [110]. (b) Temperature dependence of the longitudinal magnetoresistance MR of PrV$_2$Al$_{20}$. The solid line marks the MR = 0 position. Inset of (b): The MR develops a negative minimum, as marked by the arrows, and changes sign from negative to positive at $T_{{\rm MR}=0}\sim 8$ K. (c) Temperature dependence of $\rho$ for LaV$_2$Al$_{20}$. (d) The Kohler's scaling plot for LaV$_2$Al$_{20}$.}
\label{resistivity}
\end{figure}

In the main panel of Fig. 1(a) (right axis), we present the $f$-electron contribution to the resistivity $\rho_{4f}(T)$ for PrV$_2$Al$_{20}$ measured at various magnetic fields.  The $\rho_{4f}(T)$ component is obtained by subtracting $\rho(T)$ of LaV$_2$Al$_{20}$ at each field (Fig. 1(c)) from the raw data of PrV$_2$Al$_{20}$ (Fig. 1(a), left axis). Upon cooling, $\rho_{4f}(T)$ shows a logarithmic temperature dependence and forms a broad peak at $T_{\rm peak}\sim 40$ K, owing to the magnetic Kondo effect induced by the excited triplet states. The magnitude of $T_{\rm peak}$ represents the crystal field splitting $\Delta$ between the first excited $\Gamma_4$ triplet and the nonmagnetic $\Gamma_3$ ground-state doublet, as discussed in our previous studies \cite{Sakai2011}. In applied magnetic fields, $\rho_{4f}$ decreases in the temperature range of 8 K $\lesssim T \lesssim$ 40 K, and then begins to rise below $\sim$ 8 K. Such feature is particularly evident in the longitudinal magnetoresistivity ${\rm MR}\equiv (\rho(B,T)-\rho(0,T))/\rho(0,T)$ shown in Fig. 1(b). For \textit{B} $>$ 2T, MR develops a negative minimum at $T_{\rm min}\sim$ 15 K, as indicated by the arrows in the inset of Fig. 1(b). This negative MR of PrV$_2$Al$_{20}$ contrasts strikingly with that of typical nonmagnetic metals, in which the MR is commonly positive. Indeed, non-$f$ analogue LaV$_2$Al$_{20}$ shows positive MR in the entire measured temperature range, and obeys the Kohler's rule \cite{Kohler}, as plotted in Fig. 1 (c) and (d), respectively. These features indicate that a single scattering mechanism, probably electron-phonon scattering, is dominant in LaV$_2$Al$_{20}$. Moreover, the MR of PrV$_2$Al$_{20}$ undergoes a sign change at $T_{{\rm MR}=0}\sim 8$ K, and increases to positive values on further cooling. Such a behavior, namely negative MR with a minimum followed by a sign change at a lower temperature, closely resembles that of Ce-based heavy fermion compounds such as CeB$_6$\cite{Sumiyama1986}. In the case of Ce-based systems, this behavior is interpreted by a crossover from an incoherent metal to a coherent heavy fermion liquid with local $\textit{f}$-moments incorporated into the Fermi volume\cite{KawakamiOkiji1986}.

\begin{figure}[t]
\begin{center}
\includegraphics[keepaspectratio, scale=0.5]{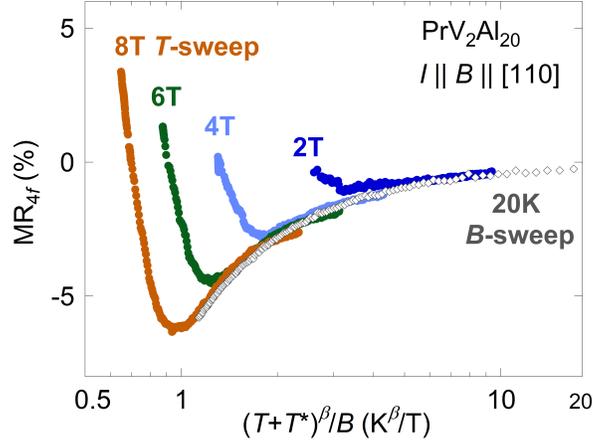}
\end{center}
\caption{(Color online)   Generalized  Schlottmann's scaling relation for $4f$ electron contribution to the MR in PrV$_2$Al$_{20}$. The filled and open symbols represent data obtained from temperature and field scans, respectively. The best scaling gives a characteristic temperature of  $T^*=9.5$ K and a power-law exponent of $\beta =0.65$ (main text).
}
\end{figure}

In conventional magnetic Kondo systems, the MR curves measured at various magnetic fields follow the Schlottmann's scaling such that the $T$ and $B$ dependences of MR appear only through the ratio $(T+T^*)/B$; the characteristic temperature $T^*$  plays the role of the single-impurity Kondo temperature\cite{Schlottmann1983,Kaczorowski2000, Batlogg1987}. Motivated by the experiment on UBe$_{13}$ \cite{Andraka1994}, theoretical work on the two-channel Kondo lattice \cite{Anders1997} provides a different scaling with a smaller exponent, namely, $\sim(T+T^*)^\beta/B$ with $\beta \sim 0.5$. Therefore, we analyze the 4$f$-electron contribution to the magnetoresistivity, MR$_{4f}$ of PrV$_2$Al$_{20}$ using a generalized form of the Schlottmann's relation
\begin{equation}
{\rm MR}_{4f}\equiv \frac{(\rho_{4f}(B,T)-\rho_{4f}(0,T))}{\rho_{4f}(0,T)}=F \left(\frac{B}{(T+T^*)^\beta}\right),
\label{eq.Anders}
\end{equation}
where ${\rm MR}_{4f}$ and $F(x)$ are the $4f$-electron component of the MR and a scaling function, respectively. Figure 2 (c) displays the scaling plot according to Eq. (1) for PrV$_2$Al$_{20}$. The scaling relation holds well for $T\gtrsim$ 10 K, where the negative MR is observed. The best scaling yields $T^* =9.5 \pm 1$ K and $\beta = 0.65 \pm 0.1$. This $\beta$ value deviates from $\beta \sim 1$ expected for the magnetic Kondo systems\cite{Andraka1995, Malinowski2005, Maple2006}; yet it is nearly identical to that found in UBe$_{13}$ ($\beta \sim 0.6$) \cite{Andraka1994}, and is consistent with the theoretically prediction for the two-channel Kondo model \cite{Anders1997}. This finding suggests the presence of the quadrupolar Kondo effect in
PrV$_2$Al$_{20}$.

To further explore the quadrupolar Kondo physics in PrV$_2$Al$_{20}$, we turn our focus to the non-Fermi liquid behavior of $\rho_{4f}(T)$ in the paraquadrupolar state. For $T_{Q}< T \lesssim$ 20 K, $\rho_{4f}(T) \sim \sqrt{T}$. To clearly illustrate this $\sqrt{T}$ dependence of $\rho_{4f}(T)$, we plot $\rho_{4f}/(\rho_0+A\sqrt{T})$  as a function of $T$ in the inset of Fig. 3(a), where $\rho_0$ is the residual resistivity and $A$ is a coefficient. 
The arrows mark the characteristic temperature $T_0$ at which $\rho_{4f}(T)$ deviates from the $\sqrt{T}$ behavior, in other words, the upper limit of $\rho/(\rho_0+A\sqrt{T})\sim 1$. The resulting $T_0$ increases slightly with field, as plotted in Fig. 4. The main panel of Fig. 3(a) shows  $\rho_{4f}(T,B)/\rho(T_0,B)$ vs $T/T_0$ for PrV$_2$Al$_{20}$. By using $T_0$, normalized resistivity $\rho_{4f}(T,B)/\rho(T=T_0,B)$ collapse onto a single curve, suggesting that $\rho_{4f}(T)$ follows a universal scaling relation, similar to the case of PrIr$_2$Zn$_{20}$ and PrRh$_2$Zn$_{20}$\cite{Onimaru2016}. Recent theoretical studies for the two-channel Kondo lattice model yield the following relation for the temperature dependence of $\rho_{4f}(T)$
\begin{equation}
\Delta \rho_{4f}(T)=\frac{a}{1+b(T_0/T)},
\label{eq.Tsuruta}
\end{equation}
where $a$, $b$ are constant parameters\cite{Tsuruta2015}.
This theoretical prediction well reproduces the experimental $\rho_{4f}(T)$ for $ 0.5 \lesssim T/T_0 \lesssim2$. Note that this temperature range coincides with the region in which both negative MR and generalized Schlottmann's scaling are observed. 

\begin{figure}[t]
\begin{center}
\includegraphics[keepaspectratio, scale=0.6]{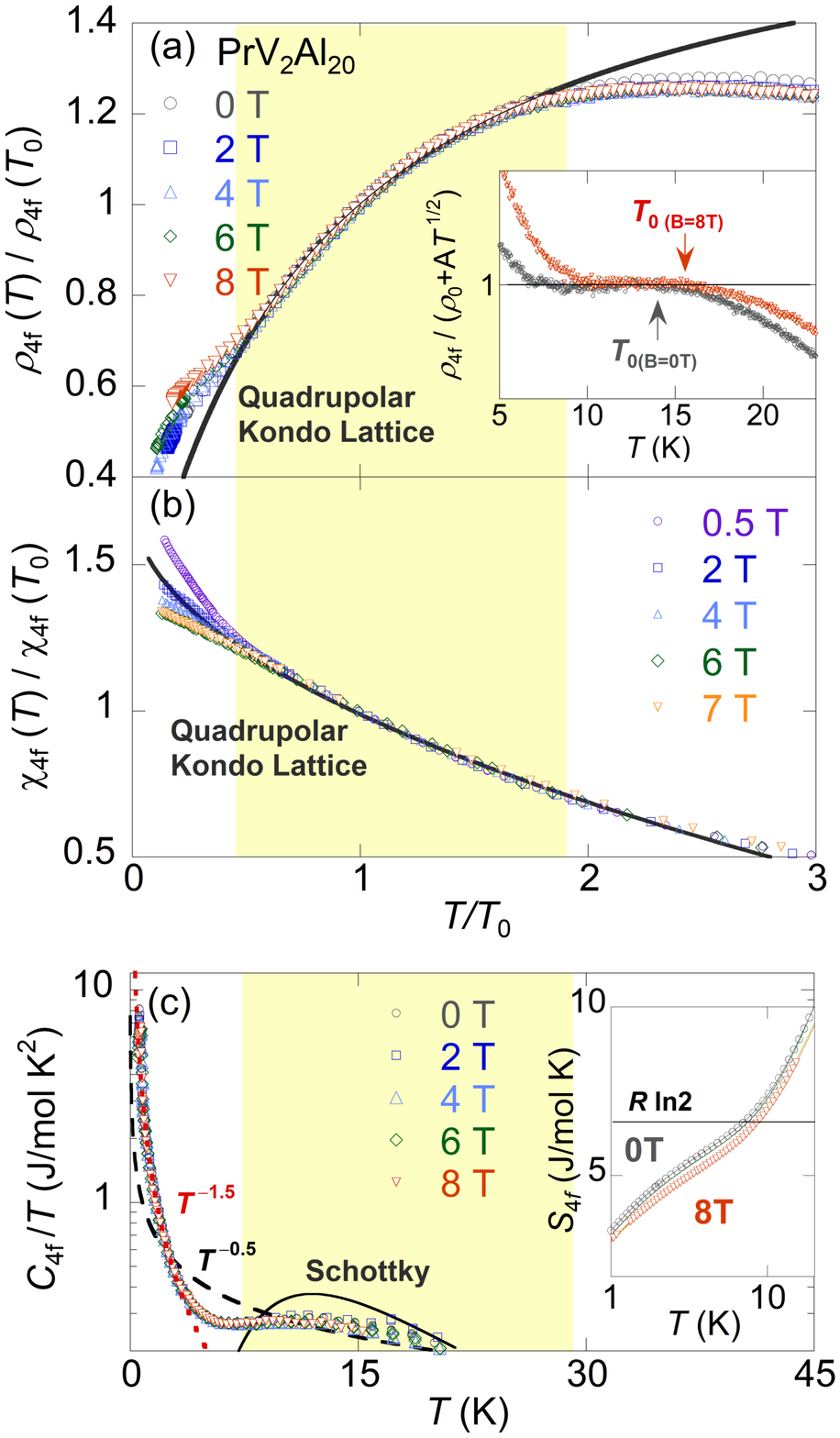}
\end{center}
\caption{(Color online) Scaling plots for (a) the 4$f$-electron contribution $\rho_{4f}$ to the electrical resistivity and (b) the 4$f$-electron contribution $\chi_{4f}$ to the magnetic susceptibility of PrV$_2$Al$_{20}$ under a [110] magnetic field. The solid lines represent the universal scaling curves obtained in the quadrupolar Kondo lattice model \cite{Tsuruta2015}. The inset of (a) shows the $T$ dependence of $\rho/(\rho_0+A\sqrt{T})$ measured at 0 and 8T. The arrows indicate the characteristic temperature $T_0$, where $\rho(T)$ begins to deviate from the $\sqrt{T}$ behavior. (c) Temperature dependence of $C_{4f}/T$ of PrV$_2$Al$_{20}$ in $B \parallel$ [110]. The solid line represents the calculated Schottky anomaly assuming that the first CEF excited state is located at 40 K above the ground-state doublet. The dotted and dashed lines represent the $T^{-1/2}$ and $T^{-3/2}$ dependence, respectively. The inset in (c) shows the $T$ dependence of $4f$ electron contribution to the entropy $S_{4f}$ at the 0 and 8 T. In all three panels, the shaded area corresponds to the region in which the quadrupolar Kondo scaling is found.}
\end{figure}

Similar to $\rho_{4f}(T)$, the normalized magnetic susceptibility $\chi_{4f}(T)/\chi_{4f}(T_0)$ can be scaled onto a single curve over roughly the same temperature range, $ 0.5 \lesssim T/T_0 \lesssim2$, as shown in Fig. 3(b). Here, we use the same $T_0$ as that for the resistivity scaling discussed above. Moreover, in this temperature regime, $\chi(T)$ is well described by the theoretical form $\chi=c-d\sqrt{T}$ for the quadrupolar Kondo system\cite{Tsuruta2015}. The scaling behavior in $\rho_{4f}(T)$ and $\chi_{4f}(T)$ again designates the quadrupolar Kondo effect as the dominant mechanism at play. The value of $T_0\sim 15$K gives the energy scale of the hybridization between the 4$f$ quadrupole moments and the conduction electrons, which is about one order of magnitude larger than that in PrIr$_2$Zn$_{20}$ and PrRh$_2$Zn$_{20}$\cite{Onimaru2016}. 

The quadrupolar Kondo effect may also manifest itself in the electronic contribution to the specific heat \cite{Tsuruta2015}. Figure 3(c) shows the $4f$ electron contribution to the specific heat divided by $T$, $C_{4f}/T$, of PrV$_2$Al$_{20}$, which is obtained after subtracting $C/T$ of LaV$_2$Al$_{20}$. The shaded area corresponds to the region where the quadrupolar Kondo scaling holds in $\rho_{4f}$ and $\chi_{4f}$. Unlike the case for $\rho$ and $\chi$, the scaling analysis of $C_{4f}/T$ is hindered most likely by the substantial continuation of CEF effect, namely, the Schottky anomaly centred at $T_{\rm Schottky}\sim 12$ K. As a result of the significant $c$-$f$ hybridization, the observed Schottky anomaly is much broader compared to the one calculated using the two-level model, assuming a doublet ground state and a triplet excited state located at $\Delta = 40$K. The CEF contribution is also reflected in the $4f$-electron contribution to the entropy, $S_{4f}$, which reaches $\sim R\ln 2$ at $T_{\rm MR=0}\sim 8$ K (inset of Fig. 3(c)); if $S_{4f}$ is governed solely by the quadrupolar Kondo coupling, this ground-state entropy of $R\ln 2$ would be released at the higher characteristic temperature of $T_0\sim 15$K.  At lower temperatures, $C_{4f}/T$ strongly increases as $\sim T^{-3/2}$. This power-law enhancement is more singular than the $\sim T^{-1/2}$ behavior predicted by the two-channel Kondo lattice model\cite{Tsuruta2015}. 

\begin{figure}[b]
\begin{center}
\includegraphics[keepaspectratio, scale=0.35]{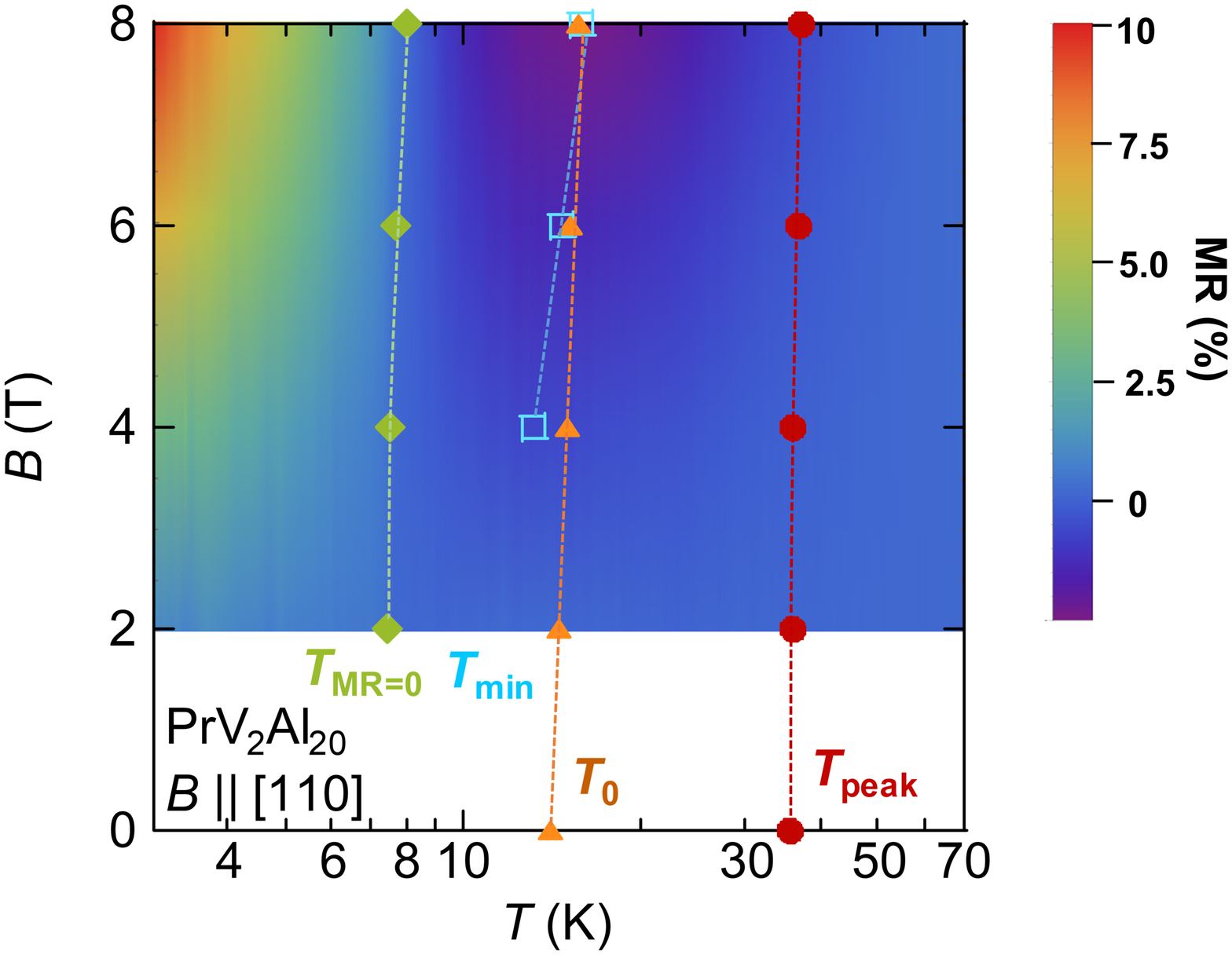}
\end{center}
\caption{(Color online) Contour plot of the magnetoresistance MR in a $B$-$T$ phase diagram of PrV$_2$Al$_{20}$ under [110] magnetic field. The characteristic temperature  $T_0$ is obtained from the universal scaling relation. $T_{\rm peak}$ and $T_{\rm min}$ mark the peak and negative minimum positions of MR, respectively. The sign change in MR takes place at $T_{\rm MR=0}$. The dashed lines are guide for the eyes.}
\end{figure}

Figure 4 displays the $B$-$T$ phase diagram of PrV$_2$Al$_{20}$, with the color code representing the magnitude of the magnetoresistance MR. The characteristic temperatures $T_0$, $T_{\rm peak}$, $T_{\rm min}$, and $T_{\rm MR=0}$ obtained from the MR curves are also plotted. As clearly demonstrated by this phase diagram, the parameter regime with valid quadrupolar Kondo scaling relations corresponds fully to that of the negative MR. 

Our results bear important implication for the interplay between the conventional magnetic Kondo effect and the quadrupolar Kondo effect in PrV$_2$Al$_{20}$. In the high temperature regime $T\gtrsim 40$ K $\sim\Delta$, the magnetic Kondo effect driven by the CEF excited triplets plays a dominant role, as evident from the logarithmic behavior of  $\rho_{4f} \sim - \ln T$, as shown in Fig. 1(a). On cooling, the nonmagnetic $\Gamma_3$ ground-state doublet becomes more occupied, whereas the population of the excited states declines. Owing to this crossover of the CEF states, the quadrupolar Kondo effect predominates over the magnetic Kondo effect at lower temperatures. This crossover from the magnetic to quadrupolar Kondo effect serves as the driving force behind the negative MR, the NFL behavior and the universal scaling relation in the temperature range of 8 K $< T <$ 30 K.

Below $T_{\rm MR=0}\sim 8$ K, the behavior of $\rho_{4f}$ and $\chi_{4f}$ deviates from the scaling relations for the quadrupolar Kondo lattice model, as shown in Fig. 3 (a) and (b), implying that another mechanism comes into play that is not taken into account in the model. Moreover, the MR develops a positive upturn for $T < T_{\rm MR=0}\sim 8$ K.  In magnetic heavy fermion systems, such positive MR is considered a signature of the Kondo lattice coherence \cite{KawakamiOkiji1986}. Correspondingly, the quadrupolar moments may couple coherently with conduction electrons at low temperatures and become a part of the Fermi volume. The quadrupole-driven coherence effect may render a new type of electronic order, such as a composite electronic order featuring symmetry breaking between the two channels in the quadrupolar Kondo lattice model\cite{Hoshino2011,Premala2015}. Another possible mechanism is the enhancement and critical slowing down of quadrupolar quantum fluctuations as a precursor of the long-range quadrupole order. In both scenarios, a significant enhancement of $C/T$ is expected to take place below $T_{\rm MR=0}$, which is compatible with the experimental observation shown in Figure. 3(c). 

The $\Gamma_3$ ground-state doublet of PrV$_2$Al$_{20}$ possesses both quadrupolar and octupolar moments, yet the latter is not considered in the quadrupolar Kondo lattice model \cite{Tsuruta2015}. Recent theoretical work on a generalized multipolar single-impurity Kondo model proposed that, by including the octupolar moment into the Kondo coupling, a new type of NFL phase may appear, which connects with the one found in the quadrupolar Kondo lattice model through a low-temperature crossover\cite{Adarsh2019}. Further theoretical studies on generalized multipolar Kondo lattice is then highly desired to allow direct comparison with our experimental results.

In summary, we have conducted magnetoresistance, magnetic susceptibility and specific heat measurements of PrV$_2$Al$_{20}$ under a [110] magnetic field. Upon cooling below $\sim 30$ K, we observed a universal scaling behavior in both $\rho_{4f}$ and $\chi_{4f}$ as expected from the quadrupolar Kondo lattice model, indicating that the quadrupolar Kondo effect serves as the primary mechanism for the NFL behavior observed in PrV$_{2}$Al$_{20}$. This finding establishes the cubic non-Kramers doublet compounds Pr$Tr_2$Al$_{20}$ ($Tr$ : Ti, V) as an essential testing ground for the theories of multipolar Kondo effect. The scaling relation is violated at low temperatures of $T\lesssim 8$ K, accompanied by a sign change from negative to positive in the MR. Such behavior might signify the onset of a new type of electronic order owing to the coherence effect in the quadrupolar Kondo lattice. Another possibility is the critical effect of quantum quadrupolar fluctuations. Further studies of the Pr$Tr_2$Al$_{20}$ ($Tr$ : Ti, V) family at ultralow temperatures and in higher magnetic fields may enable us to fully characterize the electronic phase diagram and to reveal the fate of the NFL phase and the heavy fermion superconductivity upon entering the quantum critical regime governed solely by the orbital degrees of freedom. 

\acknowledgments
{We thank Y. Shimura, H. Sato, T. Sakakibara, H. Takigawa, T. Shibauchi, Y. Tokunaga, for insightful discussions. This work is partially supported by CREST(JPMJCR18T3), Japan Science and Technology Agency, by Grants-in-Aids for Scientific Research on Innovative Areas (15H05882 and 15H05883) from the Ministry of Education, Culture, Sports, Science, and Technology of Japan, and by Grants-in-Aid for Scientific Research (19H00650) from the Japanese Society for the Promotion of Science (JSPS). The work at IQM was supported by the US Department of Energy, office of Basic Energy Sciences, Division of Materials Sciences and Engineering under grant DE-FG02-08ER46544. M. F. acknowledges support from the Japan Society for the Promotion of Science (Postdoctoral Fellowship for Research in Japan (Short-term)). M. T. was supported by Japan Society for the Promotion of Science through Program for Leading Graduate Schools (MERIT).}


\bibliographystyle{jpsj} 
\bibliography{MR}

\begin{thebibliography}{10}

\bibitem{Keimer2015}
B.~Keimer, S.~A. Kivelson, M.~R. Norman, S.~Uchida, and J.~Zaanen: Nature
  {\bfseries 518} (2015) 179.

\bibitem{Si2016}
Q.~Si, R.~Yu, and E.~Abrahams: Nat. Rev. Mater. {\bfseries 1} (2016) 16017.

\bibitem{Stewart2001}
G.~R. Stewart: Rev. Mod. Phys. {\bfseries 73} (2001) 797.

\bibitem{Gegenwart2008}
P.~Gegenwart, Q.~Si, and F.~Steglich: Nat. Phys. {\bfseries 4} (2008) 186.

\bibitem{Lohneysen2007}
H.~V. L{\"{o}}hneysen, A.~Rosch, M.~Vojta, and P.~W{\"{o}}lfle: Rev. Mod. Phys.
  {\bfseries 79} (2007) 1015.

\bibitem{Yuan2104}
H.~Q. Yuan, F.~M. Grosche, M.~Deppe, C.~Geibel, G.~Sparn, and F.~Steglich:
  Science {\bfseries 302} (2003) 2104.

\bibitem{Custers2003}
J.~Custers, P.~Gegenwart, H.~Wilhelm, K.~Neumaier, Y.~Tokiwa, O.~Trovarelli,
  C.~Geibel, F.~Steglich, C.~P{\'e}pin, and P.~Coleman: Nature {\bfseries 424}
  (2003) 524.

\bibitem{Nakatsuji2008}
S.~Nakatsuji, K.~Kuga, Y.~Machida, T.~Tayama, T.~Sakakibara, Y.~Karaki,
  H.~Ishimoto, S.~Yonezawa, Y.~Maeno, E.~Pearson, G.~G. Lonzarich, L.~Balicas,
  H.~Lee, and Z.~Fisk: Nat. Phys. {\bfseries 4} (2008) 603.

\bibitem{Matsumoto2011}
Y.~Matsumoto, S.~Nakatsuji, K.~Kuga, Y.~Karaki, N.~Horie, Y.~Shimura,
  T.~Sakakibara, A.~H. Nevidomskyy, and P.~Coleman: Science {\bfseries 331}
  (2011) 316.

\bibitem{Tomita506}
T.~Tomita, K.~Kuga, Y.~Uwatoko, P.~Coleman, and S.~Nakatsuji: Science
  {\bfseries 349} (2015) 506.

\bibitem{Coleman_2001}
P.~Coleman, C.~P{\'{e}}pin, Q.~Si, and R.~Ramazashvili: Journal of Physics:
  Condensed Matter {\bfseries 13} (2001) R723.

\bibitem{KuramotoReview2009}
Y.~Kuramoto, H.~Kusunose, and A.~Kiss: Journal of the Physical Society of Japan
  {\bfseries 78} (2009) 072001.

\bibitem{Sakai2011}
A.~Sakai and S.~Nakatsuji: J. Phys. Soc. Jpn. {\bfseries 80} (2011) 063701.

\bibitem{OnimaruReview2016}
T.~Onimaru and H.~Kusunose: Journal of the Physical Society of Japan {\bfseries
  85} (2016) 082002.

\bibitem{Cox1987}
D.~L. Cox: Phys. Rev. Lett. {\bfseries 59} (1987) 1240.

\bibitem{Cox1998}
D.~L. Cox and A.~Zawadowski: Adv. Phys. {\bfseries 47} (1998) 599.

\bibitem{Tsuruta2015}
A.~Tsuruta and K.~Miyake: J. Phys. Soc. Jpn. {\bfseries 84} (2015) 114714.

\bibitem{Adarsh2019}
A.~S. {Patri}, I.~{Khait}, and Y.~B. {Kim}: arXiv e-prints  (2019)
  arXiv:1904.02717.

\bibitem{Kangas2012}
M.~J. Kangas, D.~C. Schmitt, A.~Sakai, S.~Nakatsuji, and J.~Y. Chan: J. Solid
  State Chem. {\bfseries 196} (2012) 274.

\bibitem{Matsunami2011}
M.~Matsunami, M.~Taguchi, A.~Chainani, R.~Eguchi, M.~Oura, A.~Sakai,
  S.~Nakatsuji, and S.~Shin: Phys. Rev. B {\bfseries 84} (2011) 193101.

\bibitem{Matsubayashi2014}
K.~Matsubayashi, T.~Tanaka, A.~Sakai, S.~Nakatsuji, Y.~Kubo, and Y.~Uwatoko:
  Phys. Rev. Lett. {\bfseries 109} (2012) 187004.

\bibitem{Tsujimoto2014a}
M.~Tsujimoto, Y.~Matsumoto, T.~Tomita, A.~Sakai, and S.~Nakatsuji: Phys. Rev.
  Lett. {\bfseries 113} (2014) 267001.

\bibitem{Sato2012}
T.~J. Sato, S.~Ibuka, Y.~Nambu, T.~Yamazaki, T.~Hong, A.~Sakai, and
  S.~Nakatsuji: Phys. Rev. B {\bfseries 86} (2012) 184419.

\bibitem{Ito2011}
T.~U.~Ito, W.~Higemoto, K.~Ninomiya, H.~Luetkens, C.~Baines, A.~Sakai, and
  S.~Nakatsuji: J. Phys. Soc. Jpn. {\bfseries 80} (2011) 113703.

\bibitem{Ito2015}
T.~U. Ito, W.~Higemoto, A.~Sakai, M.~Tsujimoto, and S.~Nakatsuji: Phys. Rev. B
  {\bfseries 92} (2015) 125151.

\bibitem{Taniguchi2016}
T.~Taniguchi, M.~Yoshida, H.~Takeda, M.~Takigawa, M.~Tsujimoto, A.~Sakai,
  Y.~Matsumoto, and S.~Nakatsuji: J. Phys. Soc. Jpn. {\bfseries 85} (2016)
  113703.

\bibitem{Nakanishi2018}
Y.~Nakanishi, M.~Taniguchi, M.~Nakamura, J.~Hasegawa, R.~Ohyama, M.~Nakamura,
  M.~Yoshizawa, M.~Tsujimoto, and S.~Nakatsuji: Physica B: Cond. Matt.
  {\bfseries 536} (2018) 125 .

\bibitem{Tokunaga2013}
Y.~Tokunaga, H.~Sakai, S.~Kambe, A.~Sakai, S.~Nakatsuji, and H.~Harima: Phys.
  Rev. B {\bfseries 88} (2013) 085124.

\bibitem{Machida2015}
Y.~Machida, T.~Yoshida, T.~Ikeura, K.~Izawa, A.~Nakama, R.~Higashinaka,
  Y.~Aoki, H.~Sato, A.~Sakai, S.~Nakatsuji, N.~Nagasawa, K.~Matsumoto,
  T.~Onimaru, and T.~Takabatake: J. Phys.: Conference Series {\bfseries 592}
  (2015) 012025.

\bibitem{Shimura2015}
Y.~Shimura, M.~Tsujimoto, B.~Zeng, L.~Balicas, A.~Sakai, and S.~Nakatsuji:
  Phys. Rev. B {\bfseries 91} (2015) 241102.

\bibitem{Anders1997}
F.~B. Anders, M.~Jarrell, and D.~L. Cox: Phys. Rev. Lett. {\bfseries 78} (1997)
  2000.

\bibitem{Kohler}
A.~B. Pippard: Cambridge University Press, Cambridge  (1989).

\bibitem{Sumiyama1986}
A.~Sumiyama, Y.~Oda, H.~Nagano, Y.~Onuki, K.~Shibutani, and T.~Komatsubara: J.
  Phys. Soc. Jpn. {\bfseries 55} (1986) 1294.

\bibitem{KawakamiOkiji1986}
N.~Kawakami and A.~Okiji: J. Phys. Soc. Jpn. {\bfseries 55} (1986) 2114.

\bibitem{Schlottmann1983}
P.~Schlottmann: Z. Phys. Cond. Matt. {\bfseries 51} (1983) 223.

\bibitem{Kaczorowski2000}
D.~Kaczorowski, B.~Andraka, R.~Pietri, T.~Cichorek, and V.~I. Zaremba: Phys.
  Rev. B {\bfseries 61} (2000) 15255.

\bibitem{Batlogg1987}
B.~Batlogg, D.~J. Bishop, E.~Bucher, B.~{Golding Jr}, A.~P. Ramirez, Z.~Fisk,
  J.~L. Smith, and H.~R. Ott: J. Magn. Magn. Mater {\bfseries 63-64} (1987)
  441.

\bibitem{Andraka1994}
B.~Andraka and G.~R. Stewart: Phys. Rev. B {\bfseries 49} (1994) 12359.

\bibitem{Andraka1995}
B.~Andraka: Phys. Rev. B {\bfseries 52} (1995) 16031.

\bibitem{Malinowski2005}
A.~Malinowski, M.~F. Hundley, C.~Capan, F.~Ronning, R.~Movshovich, N.~O.
  Moreno, J.~L. Sarrao, and J.~D. Thompson: Phys. Rev. B {\bfseries 72} (2005)
  184506.

\bibitem{Maple2006}
M.~B. Maple, N.~P. Butch, N.~A. Frederick, P.-C. Ho, J.~R. Jeffries, T.~A.
  Sayles, T.~Yanagisawa, W.~M. Yuhasz, S.~Chi, H.~J. Kang, J.~W. Lynn, P.~Dai,
  S.~K. McCall, M.~W. McElfresh, M.~J. Fluss, Z.~Henkie, and A.~Pietraszko:
  PNAS {\bfseries 103} (2006) 6783.

\bibitem{Onimaru2016}
T.~Onimaru, K.~Izawa, K.~T. Matsumoto, T.~Yoshida, Y.~Machida, T.~Ikeura,
  K.~Wakiya, K.~Umeo, S.~Kittaka, K.~Araki, T.~Sakakibara, and T.~Takabatake:
  Phys. Rev. B {\bfseries 94} (2016) 075134.

\bibitem{Hoshino2011}
S.~Hoshino, J.~Otsuki, and Y.~Kuramoto: Phys. Rev. Lett. {\bfseries 107} (2011)
  247202.

\bibitem{Premala2015}
P.~Chandra, P.~Coleman, and R.~Flint: Phys. Rev. B {\bfseries 91} (2015)
  205103.

\end{thebibliography}

\end{document}